\documentclass{emulateapj}

\newcommand{\msun}{M$_{\sun}$}

\newcommand{\ldl}{$\lambda/{\Delta}{\lambda}$}

\newcommand{\teff}{T$_{\rm eff}$}

\newcommand{\ki}{\ion{K}{1}}

\newcommand{\nai}{\ion{Na}{1}}

\newcommand{\meth}{CH$_4$}
\newcommand{\wat}{H$_2$O}

\newcommand{\kms}{km~s$^{-1}$}

\newcommand{\name}{WISE~J104915.57$-$531906.1}
\newcommand{\namesh}{WISE~J1049$-$5319}

\slugcomment{Submitted to ApJ 27 March 2013}

\shorttitle{Near-Infrared Spectra of WISE~J1049$-$5319AB}
\shortauthors{Burgasser et al.}

\begin{document}

\title{Resolved Near-Infrared Spectroscopy of WISE~J104915.57$-$531906.1AB: A Flux-Reversal Binary at the L dwarf/T dwarf Transition}

\author{
Adam J.\ Burgasser\altaffilmark{1},
Scott S.\ Sheppard\altaffilmark{2}, and
K.\ L.\ Luhman\altaffilmark{3,4}
}

\altaffiltext{1}{Center for Astrophysics and Space Science, University of California San Diego, La Jolla, CA 92093, USA; aburgasser@ucsd.edu}
\altaffiltext{2}{Department of Terrestrial Magnetism, Carnegie Institution of Washington, 5241 Broad Branch Rd. NW, Washington, DC 20015, USA}
\altaffiltext{3}{Department of Astronomy and Astrophysics, The Pennsylvania State University, University Park, PA 16802, USA}
\altaffiltext{4}{Center for Exoplanets and Habitable Worlds, The Pennsylvania State University, University Park, PA 16802, USA}

\begin{abstract}
We report resolved near-infrared spectroscopy and photometry of the recently identified brown dwarf binary 
WISE~J104915.57$-$531906.1AB, located 2.02$\pm$0.15~pc from the Sun.
Low-resolution spectral data from Magellan/FIRE and IRTF/SpeX reveal strong {\wat} and CO absorption features in the spectra of both components, with the secondary also exhibiting weak CH$_4$ absorption at 1.6~$\micron$ and 2.2~$\micron$.  Spectral indices and comparison to low-resolution spectral standards indicate component  types of L7.5 and T0.5, the former consistent with the optical classification of the primary.
Relative photometry reveals a flux reversal between the $J$- and $K$-bands, with the T dwarf component being brighter in the 0.95--1.3~$\micron$ range.  As with other L/T transition binaries, this reversal likely reflects significant depletion of condensate opacity across the transition, a behavior that may be enhanced in WISE~J1049$-$5319AB if the unusual red color of its L dwarf component is indicative of thick clouds.  
On the other hand, differing cloud properties may have modified the evolutionary paths of these two components, and we propose a scenario in which the cooler secondary could be the more massive of the two components.
Fortunately, the proximity, brightness and small separation (3.12$\pm$0.25~AU) of this system make it amenable to astrometric and radial velocity orbit measurement during its estimated 25~yr orbit, providing a rare opportunity for the direct determination of individual brown dwarf masses and a unique benchmark for studying cloud evolution across the L dwarf/T dwarf transition.
\end{abstract}


\section{Introduction}

The low temperatures and luminosities of brown dwarfs near the Sun are a consequence of their age and inability to sustain core hydrogen fusion \citep{1962AJ.....67S.579K,1963PThPh..30..460H}.  As these objects evolve down the spectral sequence (through the M, L, T and Y spectral classes; \citealt{2005ARA&A..43..195K}), their spectral energy distributions become increasingly complex as molecular compounds come to dominate photospheric opacity. 
These compounds include condensed species: minerals and metals in late-M and L dwarf atmospheres \citep{1989ApJ...338..314L,1996A&A...305L...1T}, sulfide and alkali salts in late-T and Y dwarf atmospheres \citep{2010ApJ...725.1405B,2012ApJ...756..172M}.  Grain scattering and absorption from these condensates cause significant redistribution of flux, veiling of molecular bands, changes in the gas chemistry, modification of cooling rates, and even variability if the condensates are not uniformly distributed in the photosphere \citep{2001ApJ...556..357A,2001ApJ...556..872A,1999ApJ...512..843B,2008ApJ...689.1327S,2012ApJ...750..105R}.
Models of cool atmospheres parameterizing ``cloudiness''  are now used in fits of brown dwarf spectra
\citep{2009ApJ...702..154S}, and condensate clouds are also seen as playing a major role in shaping the spectra of directly imaged giant exoplanets \citep{2011ApJ...729..128C,2011ApJ...733...65B,2012ApJ...753...14S,2012ApJ...754..135M}.

Discerning the parameters relevant to condensate formation and evolution in cool atmospheres is partly hindered by uncertainties in the characteristics of individual brown dwarfs, which for a given spectral type can span a broad range of age, mass and composition.  Resolved brown dwarf binaries have proven to be useful in this regard, as their common distance and genesis eliminates
many of the uncertainties in their (relative) physical and spatial characteristics. Such systems are also amenable to direct mass measurement (e.g., \citealt{2001ApJ...560..390L,2009ApJ...692..729D,2010ApJ...711.1087K}), allowing direct comparison to and tests of evolutionary models.  
Binaries with L dwarf and T dwarf components have been  particularly valuable in probing the disappearance of mineral clouds at this spectral transition. 
A handful are found to be ``flux-reversal'' pairs, in which the later-type secondary is brighter in the 1.0--1.3~$\micron$ region than its earlier-type (and overall more luminous) primary \citep{2006ApJS..166..585B,2006ApJ...647.1393L,2008ApJ...685.1183L}.
These binaries validate the 1~$\micron$ brightening found in near-infrared color magnitude diagrams of field L and T dwarfs (the ``J-band bump''; \citealt{2003AJ....126..975T}), and suggest that this brightening may be a general feature of brown dwarf evolution.  Physical interpretation of this brightening remains under debate, however.  It may be evidence of a rapid and/or patchy disruption of the condensate cloud layer  \citep{2002ApJ...571L.151B,2004AJ....127.3553K} or it may reflect variation among sources due to  metallicity, mass, age, rotation or other physical characteristics  \citep{2006ApJ...640.1063B}.  Unfortunately, binaries straddling the L dwarf/T dwarf transition are relatively rare, and the compact separations typical of brown dwarf pairs (1--10~AU; \citealt{2007prpl.conf..427B}) means that they must be very near to the Sun to be resolved. 

The recent discovery of the brown dwarf binary {\name}AB (hereafter {\namesh}AB; \citealt{2013ApJ...767L...1L}), at a distance of only 2.02$\pm$0.15~pc from the Sun, therefore represents an outstanding opportunity to study the L/T transition in considerable detail.
\citet{2013ApJ...767L...1L} presented an optical spectrum of {\namesh}A\footnote{We follow the convention of \citet{2013ApJ...767L...1L}, labeling the southeastern component A and the northwestern component B based on their relative $i$-band magnitudes.} consistent with a spectral type of L8$\pm$1 and revealing the presence of strong Li~I absorption, implying that this component has a mass below 0.065~{\msun} \citep{1992ApJ...389L..83R,1993ApJ...404L..17M}.  The observation that the secondary is fainter in the red optical ($\Delta{i}$ = 0.45) indicates that this component could be a late-L or T dwarf.
In this article, we report resolved near-infrared photometry and spectroscopy of the {\namesh}AB system using the Folded-Port Infrared Echellette (FIRE; \citealt{2008SPIE.7014E..27S,2010SPIE.7735E..38S}) on the Magellan 6.5m Baade Telescope at Las Campanas Observatory, and the SpeX spectrometer \citep{2003PASP..115..362R} on the 3.0m NASA Infrared Telescope Facility (IRTF). 
In Section~2 we describe our observations and reduction procedures;
in Section~3 we analyze the data, determining component spectral types and magnitudes, and identify the ``flux reversal'' nature of the system.
We discuss our results in Section~4, placing the components of {\namesh}AB in context with other late-L and T dwarfs, examining the role of cloud thickness in the flux-reversal nature and evolutionary history of this system, and motivate mass measurements based on updated predictions of its orbital properties.

\section{Observations}

\subsection{Magellan/FIRE Prism Spectroscopy and Imaging}

{\namesh} was observed with Magellan/FIRE on 12 March 2013 (UT) in photometric conditions with seeing of 0$\farcs$8 at $J$-band.  We deployed both the cross-dispersed echelle mode ({\ldl} $\approx$ 6000) and prism-dispersed mode ({\ldl} $\approx$ 300) with the 0$\arcsec$6 wide slit aligned to the parallactic angle. However, due to a slit motor error, only the prism-dispersed data were useable.  Each component was placed separately in the slit, and a series of 1~s exposures (10 for the SE component, 8 for the NW component) were obtained for each in an ABBA dither pattern, nodding 9$\arcsec$ along the 30$\arcsec$ slit.  We also observed the A0~V star HD~99338 ($V$ = 8.26) in six 1~s nodded exposures.  NeAr and quartz lamp exposures, reflected off 
of the Baade secondary screen, were obtained with the target and A0~V
stellar observations for wavelength and pixel response calibration, respectively.  Data were reduced using the FIREHOSE low-dispersion reduction package ({\it firehose\_ld}), which produces a 2D estimate of the sky spectrum to remove the background in each exposure \citep{2003PASP..115..688K}, determines a 2D wavelength mapping from the NeAr arc exposure, and extracts 1D source and A0~V spectra and variances.  Modified routines from the SpeXtool package \citep{2004PASP..116..362C} were  used to scale and combine the individual spectra, correct for telluric absorption and apply a relative flux calibration following the procedures described in \citet{2003PASP..115..389V}.
Signals-to-noise (S/N) for both components peaked at $\approx$400 in the $K$-band region.

\begin{figure*}
\center
\includegraphics[width=3in]{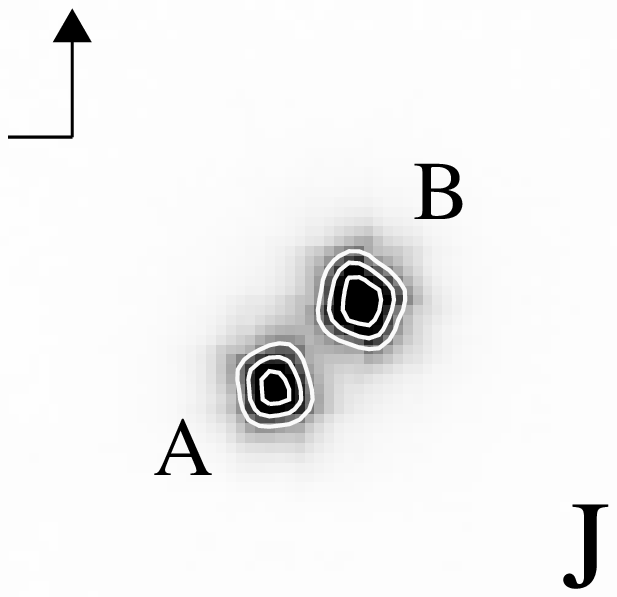}
\includegraphics[width=3in]{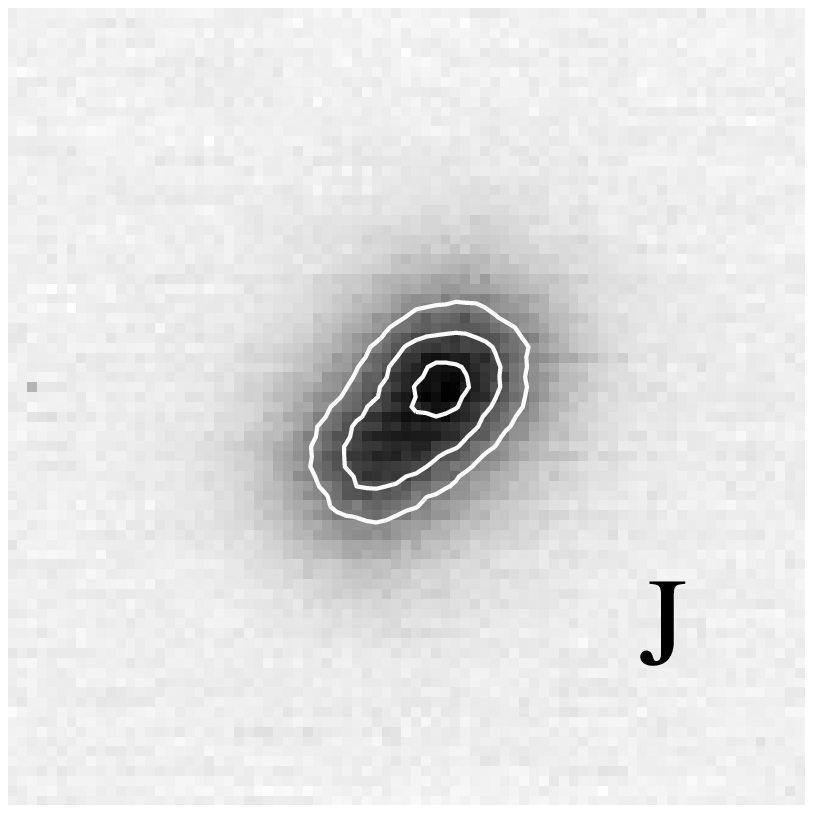}
\includegraphics[width=3in]{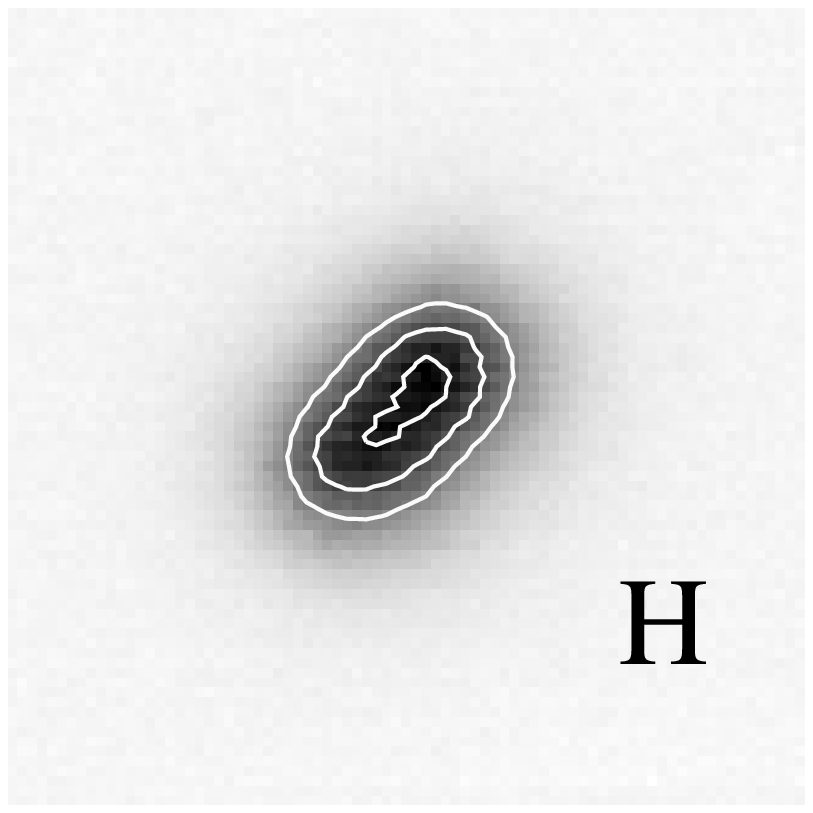}
\includegraphics[width=3in]{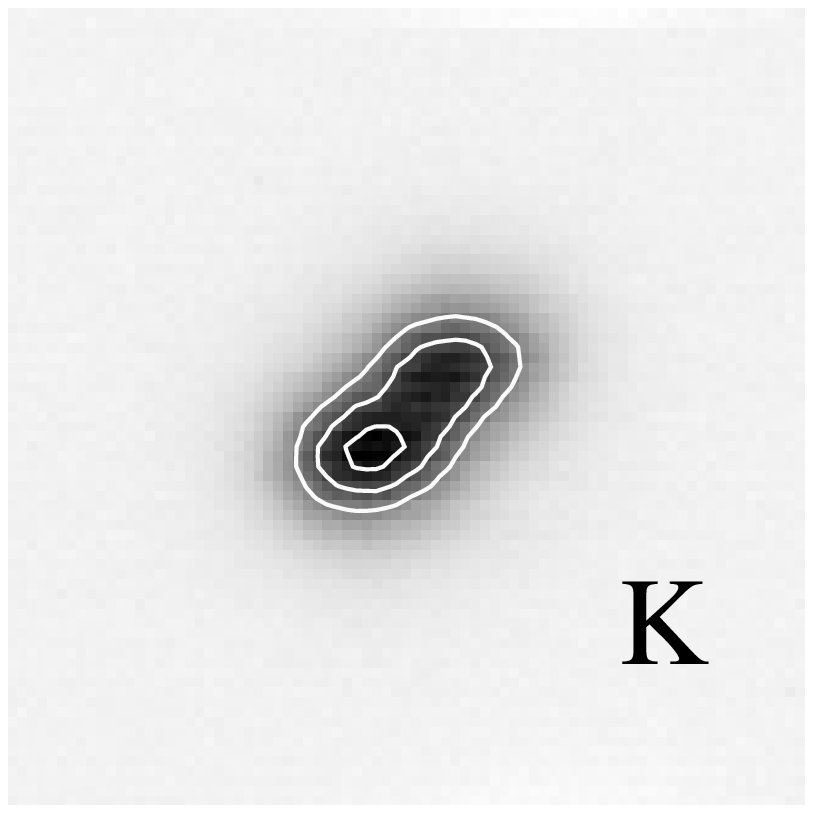}
\caption{FIRE (upper left frame) and SpeX slitviewer images of the {\namesh}AB pair in MKO $J$, $H$ and $K$-bands (labeled).
All three images are oriented with North up and East to the left, and display 9$\farcs$7$\times$9$\farcs$7 fields-of-view. Contours are indicated at 50\%, 75\% and 95\% peak flux in the FIRE image, and 50\%, 70\% and 90\% peak flux in the SpeX images.  The individual components are labeled in the FIRE image.
}
\label{fig:images}
\end{figure*}

We also obtained a resolved image of the binary using FIRE's acquisition camera, which has a 50$\arcsec$$\times$50$\arcsec$ field of view (FOV), a pixel scale of 0$\farcs$147, and a fixed MKO\footnote{Mauna Kea Observatory filter system; see \citet{2002PASP..114..180T} and \citet{2002PASP..114..169S}.} $J$-band filter.  Several 1~s exposures were obtained with the binary on and off slit; these were median-combined to produce an overall sky frame that was subtracted from all images. We extracted a 9$\farcs$7$\times$9$\farcs$7 (49$\times$49~pixels) subframe from a single image in which the binary is well-separated from the slit, shown in Figure~\ref{fig:images} and analyzed below.

\subsection{IRTF/SpeX Spectroscopy and Imaging}

{\namesh} was observed with IRTF/SpeX on 15 March 2013 (UT) in cloudy and windy conditions with poor and variable seeing (1$\farcs$5-2$\farcs$0 at $J$-band).  We deployed the 0$\farcs$5 slit and prism-dispersed mode to obtain {\ldl} $\approx$ 120 spectra covering 0.7--2.5~$\micron$.  In this case, the slit was aligned along the binary axis (position angle of 5$\degr$, 50$\degr$ off parallactic) to obtain concurrent spectra, and six 90~s exposures were obtained in an ABBA dither pattern at an airmass of 3.3--3.4.  
We also observed the A0 V star HD~92518 ($V$ = 6.87) at an airmass of 3.3
with the slit aligned to the same position angle.
Internal flat field and Ar arc lamp exposures were obtained 
for pixel response and wavelength calibration.
Data were reduced using SpeXtool, applying standard settings.
Due to the poor seeing, we did not attempt to extract component spectra; rather, we extracted the combined-light spectrum of the binary using a wide aperture.  
Average S/N was roughly 400 in the $J$, $H$ and $K$-band peaks, respectively. 
While combining the individual spectral frames, we verified that the variable cloud extinction 
during the observation was grey and had minimal impact on the observed spectral shape;
however, differential color refraction may still be an issue, and is addressed below.

We obtained images of the binary on the same night using the SpeX slit-viewing camera (60$\arcsec$$\times$60$\arcsec$ FOV, pixel scale 0$\farcs$12)
in each of the MKO $J$, $H$ and $K$ filters, with the instrument oriented at a position angle of 0$\degr$.  Four exposures were obtained in each filter using a two-point dither pattern with a 7$\arcsec$ nod, with total integrations of 32~s, 56~s and 84~s, respectively.  We interleaved these with observations of a nearby red star 2MASS~J10490107$-$5317252 ($J$ = 10.75, $J-K_s$ = 1.16) for point spread function (PSF) calibration.  Frames were pair-wise subtracted to remove sky background, mirror-flipped along the $y$-axis to reproduce sky orientation, and 9$\farcs$7$\times$9$\farcs$7 (81$\times$81~pixels) subframes were excised for analysis.  The frames with the best seeing in each of the filters are shown in Figure~\ref{fig:images}.

\section{Analysis}

\subsection{Spectral Characteristics of {\namesh}}

\begin{figure*}
\epsscale{1.0}
\plotone{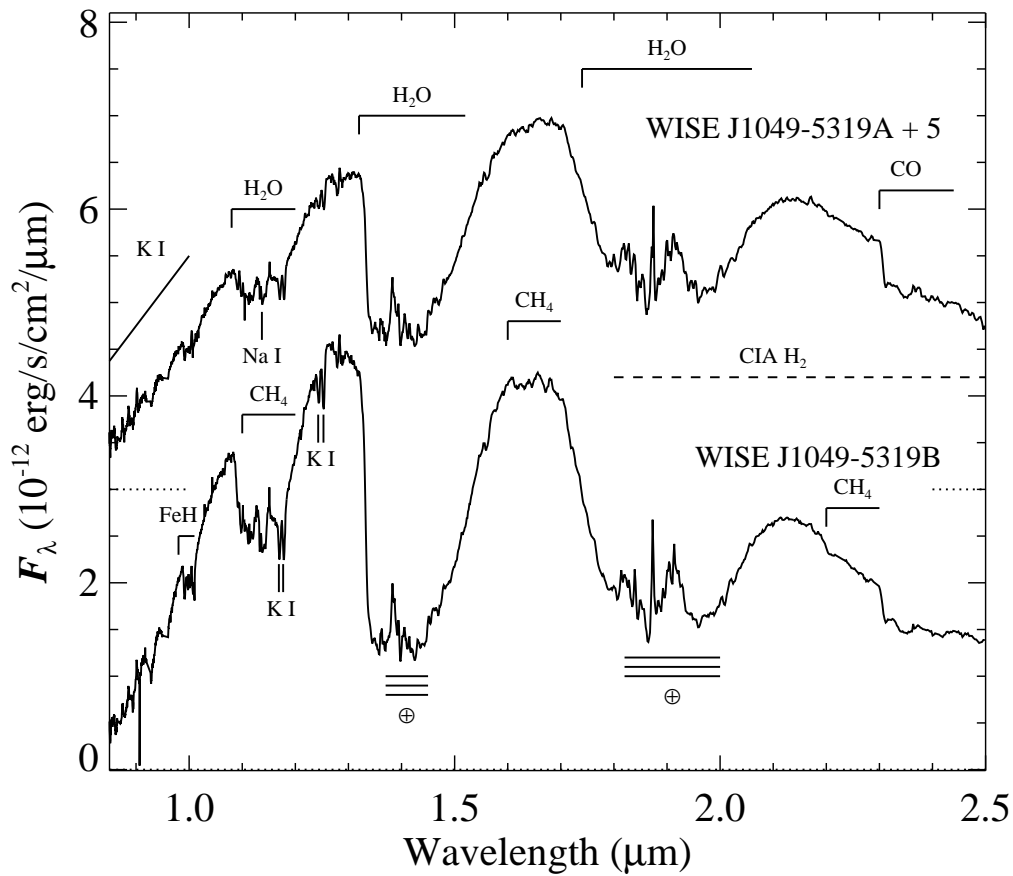}
\caption{Reduced FIRE prism spectra of {\namesh}A (top) and B (bottom), both calibrated to absolute flux units.  {\namesh}A is offset by 5$\times$10$^{-12}$ erg~s$^{-1}$~cm$^{-2}$~$\micron^{-1}$ as indicated by the dotted lines.  Major absorption features are labeled, as well as regions of strong telluric absorption ($\oplus$).
}
\label{fig:fire_prism}
\end{figure*}

The reduced FIRE spectra of the {\namesh}AB components are shown in Figure~\ref{fig:fire_prism}, scaled to their inferred absolute flux densities as described below.  Strong absorption features typical of late-L type brown dwarfs are present, notably deep {\wat} bands  at 1.4 and 1.9~$\micron$;
strong CO absorption at 2.3~$\micron$; marginally resolved {\nai} and {\ki} doublets at 1.14, 1.17 and 1.25~$\micron$;
and a steep 0.8--1.1~$\micron$ spectral slope, shaped primarily by the pressure-broadened red wing
of the 0.77~$\micron$ {\ki} doublet \citep{2000ApJ...531..438B}.
{\namesh}B also exhibits weak absorption from {\meth} at 1.6~$\micron$ and  2.2~$\micron$,
characteristic of an early-type T dwarf \citep{2006ApJ...637.1067B}. 
Enhanced absorption at 1.15~$\micron$ can also be attributed in part to {\meth} absorption.  
The near-infrared {\ki} lines are stronger 
in the spectrum of this component, and a hint of FeH can be seen at 
0.99~$\micron$ (the Wing-Ford band).
The overall near-infrared spectral energy distribution of {\namesh}B is bluer than that of {\namesh}A.

\begin{figure}
\epsscale{0.8}
\plotone{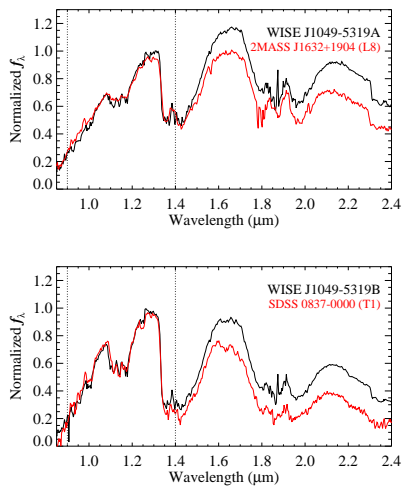}
\caption{Comparison of the FIRE prism spectra of {\namesh}A (top) and B (bottom, in black) to best match near-infrared spectral standards 2MASSW J1632291+190441 (L8) and  SDSSp J083717.22-000018.3 (T1, in red), following the method of \citet{2010ApJS..190..100K}.  All spectra are normalized.  The comparison region of 0.9--1.4~$\micron$ is indicated; both sources are notably redder than their comparison stars.  SpeX data for the standards are from \citet{2006ApJ...637.1067B} and \citet{2007ApJ...659..655B}.
}
\label{fig:standards}
\end{figure}

Spectral classifications were determined through spectral indices and comparison to spectral standards.  We used the indices and spectral type/index relations defined by \citet{2002ApJ...564..466G, 2006ApJ...637.1067B}, and \citet{2007ApJ...659..655B}; values are listed in Table~\ref{tab:classify}.  We find mean classifications of L7.5$\pm$0.9 and T0.5$\pm$0.7, consistent with the characteristics described above; the former is also consistent with the L8$\pm$1 optical classification of {\namesh}A reported in \citet{2013ApJ...767L...1L}.
We then compared the 0.9--1.4~$\micron$ spectra to low-resolution spectral standards defined
in \citet{2010ApJS..190..100K}, following the prescription for near-infrared classification outlined in that work.  The best-match standards for {\namesh}A and B are the L8 2MASSW~J1632291+190441 \citep{1999ApJ...519..802K} and the T1 SDSSp~J083717.22$-$000018.3 \citep{2000ApJ...536L..35L}, respectively (Figure~\ref{fig:standards}); these are consistent with the index types.  Note that both components have significantly redder spectral energy distributions than their corresponding standards.
Finally, we compared the full 0.9--2.4~$\micron$ FIRE spectra to low-resolution templates from the SpeX prism Spectral Libraries\footnote{\url{http://www.browndwarfs.org/spexprism}; see \citet{2010ApJ...710.1142B} for details.}.  We found best matches to the L8 SDSSp~J085758.45+570851.4 and the L9.5 SDSSp~J083008.12+482847.4 \citep{2002ApJ...564..466G} for {\namesh}A and B, respectively, and mean near-infrared classifications of L7$\pm$0.9 and L9.5$\pm$0.5. 
Combining these analyses, we assign classifications of L7.5 and T0.5 for the two components, which places them squarely across the L/T transition.

\begin{deluxetable*}{lccccc}
\tabletypesize{\footnotesize}
\tablecaption{Classification Indices for {\name}AB\label{tab:classify}}
\tablewidth{0pt}
\tablehead{
\colhead{} &
\multicolumn{2}{c}{{\namesh}A} &
\multicolumn{2}{c}{{\namesh}B} \\
\cline{2-3} \cline{4-5}
\colhead{Index} &
\colhead{Value} &
\colhead{SpT} &
\colhead{Value} &
\colhead{SpT} &
\colhead{Ref} \\
}
\startdata
{\wat}-J & 0.672$\pm$0.010 & L8 &  0.588$\pm$0.010 & T0.5 &  1  \\
{\meth}-J & 0.870$\pm$0.010 & {\nodata} &  0.734$\pm$0.010 & {\nodata} &  1  \\
{\wat}-H & 0.681$\pm$0.010 & L8 &  0.574$\pm$0.010 & T0.5 &  1  \\
{\meth}-H & 1.093$\pm$0.010 & {\nodata} &  1.073$\pm$0.010 & T1 &  1  \\
{\wat}-K & 0.716$\pm$0.010 & L7 &  0.643$\pm$0.010 & T1 &  1  \\
{\meth}-K & 0.936$\pm$0.010 &  {\nodata} & 0.845$\pm$0.010 &  {\nodata} &  1  \\
{\wat}-1.2 & 1.519$\pm$0.010 & {\nodata} &  1.734$\pm$0.010 & T1 &  2  \\
{\wat}-1.5 & 1.751$\pm$0.010 & L8 &  2.213$\pm$0.010 & T1 &  2  \\
{\meth}-1.6 & 1.073$\pm$0.010 & {\nodata} &  1.093$\pm$0.010 & T1 &  2  \\
{\meth}-2.2 & 1.068$\pm$0.010 & L6 &  1.183$\pm$0.010 & L9 &  2  \\
K/J & 0.903$\pm$0.010 &  {\nodata} & 0.578$\pm$0.010 &  {\nodata} &  1  \\
\cline{1-6}
Index Mean & {\nodata} & L7.5$\pm$0.9  &  {\nodata} & T0.5$\pm$0.7 & {\nodata} \\
Core Spectral Match & {\nodata} & L8$\pm$0.5  &  {\nodata} & T1$\pm$0.5 & 3 \\
Full Spectral Match & {\nodata} & L7$\pm$0.9  &  {\nodata} & L9.5$\pm$0.5 & 4 \\
\cline{1-6}
Adopted Types & {\nodata} & L7.5  &  {\nodata} & T0.5 & {\nodata} \\
\enddata
\tablerefs{(1) \citet{2006ApJ...637.1067B}; (2) \citet{2002ApJ...564..466G}; (3) \citet{2010ApJS..190..100K}; (4) Based on comparison to SpeX templates and using near-infrared spectral types as computed in \citet{2007ApJ...659..655B}.}
\end{deluxetable*}


\subsection{Component Photometry: A Flux Reversal Binary}

Resolved photometry by \citet{2013ApJ...767L...1L} identified {\namesh}A as the brighter of the two sources at $i$-band, but inspection of the images in Figure~\ref{fig:images} indicates that the two components ``flip'' in relative brightness, with {\namesh}B being brighter at $J$ but fainter at $K$ ({\namesh}B appears to be marginally brighter at $H$ as well).  
To quantify the amplitude of this reversal, we performed PSF-fitting analyses on our FIRE and SpeX images using a Monte Carlo Markov Chain (MCMC) technique.  For the FIRE image, our PSF model was a 2D ellipsoidal gaussian for which the major and minor axes were allowed to vary separately in width and orientation. For the SpeX images, we used both gaussian profiles and images of the PSF star obtained in the same filter as independent checks, and found that the PSF star provided much more robust fits.  Following initial guesses for the pixel positions of both primary and secondary components and their integrated fluxes, our code explored the model parameter space (including gaussian PSF shape) in randomized steps drawn from a normal distribution, and compared model and data at each step using the $\chi^2$ statistic.  Sub-pixel shifts for the SpeX PSF model were made using a damped sinc function based on code developed by John Spencer and Mike Ressler.  
We found that chain lengths of 2000 (with a 200 step burn-in) were sufficient for convergence.
For the FIRE analysis, we marginalized the distribution of each parameter in the single chain to determine uncertainties, and included a 5\% systematic uncertainty to account for the non-gaussian PSF shape. For the SpeX analysis, we used the mean and standard deviation of the 16 binary and PSF image pairings in each filter as our overall measurement and uncertainty.  Separations and position angles from all four image sets were averaged after weighting by individual uncertainties.

\begin{figure*}
\epsscale{1}
\plotone{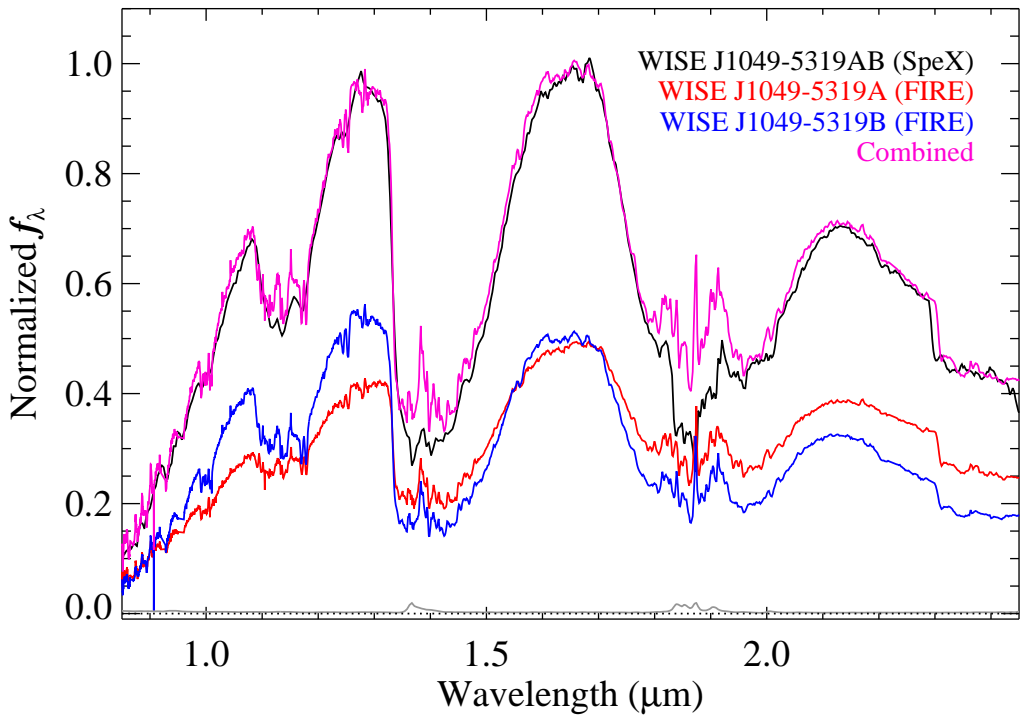}
\caption{Relative flux scaling of the FIRE spectra of {\namesh}A (red) and B (blue) based on the measured relative photometry (Table~\ref{tab:photometry}).  The summed spectrum (purple) is an good match to the combined-light SpeX spectrum of the system (black; reddened by $A_V$ = 0.6 to account for differential color refraction).  Note
that the secondary is brighter than the primary between 0.95--1.3~$\micron$ and 1.55--1.70~$\micron$.   The relative spectrophotometric magnitudes (MKO system) of the combination shown are $\Delta{J}$ = $-$0.25, $\Delta{H}$ = 0.02 and $\Delta{K}$ = 0.26.
}
\label{fig:binfit}
\end{figure*}

Results from these fits are listed in Table~\ref{tab:photometry}.  The relative photometry 
confirms the observed flux reversal:  {\namesh}B is 0.31$\pm$0.05~mag brighter at $J$ (combination of FIRE and SpeX analyses), 0.02$\pm$0.06~mag brighter at $H$ (a marginal detection) and 0.29$\pm$0.16~mag fainter at $K$.
The reversal at $J$ is highly significant, although not as extreme as that reported for the T1+T5 binary 2MASS J14044941$-$3159329AB (\citealt{2008ApJ...685.1183L,2012ApJS..201...19D}; $\Delta{J}$ = $-$0.54$\pm$0.08).  Figure~\ref{fig:binfit} displays the relative fluxes of the two components optimally scaled according to the relative magnitudes.  {\namesh}B is significantly brighter than {\namesh}A between 0.95~$\micron$ and 1.3~$\micron$, most notably at the 1.05~$\micron$ $Y$- and 1.27~$\micron$ $J$-band peaks, where the excessive flux reaches 40\%. {\namesh}B is also marginally brighter across the 1.55--1.7~$\micron$ $H$-band peak.  Importantly, these are the wavebands where condensate scattering opacity plays a prominent role in shaping brown dwarf spectra \citep{2001ApJ...556..872A}.  As shown in Figure~\ref{fig:binfit}, the sum of these scaled spectra are an excellent match to the combined light SpeX spectrum of the binary, which required modest reddening  ($A_V = 0.6$) to reproduce the combined light colors, likely to account for differential color refraction at the high airmass of the observation.

\begin{deluxetable}{lccc}
\tabletypesize{\footnotesize}
\tablecaption{Relative Photometry for {\name}AB\label{tab:photometry}}
\tablewidth{0pt}
\tablehead{
\colhead{Parameter} &
\colhead{A} &
\colhead{B} &
\colhead{$\Delta$} \\
}
\startdata
$\rho$ ($\arcsec$) & \nodata & \nodata & 1$\farcs$54$\pm$0$\farcs$04 \\	
$\rho$ (AU) & \nodata & \nodata & 3.12$\pm$0.25 \\ 
PA ($\degr$) & \nodata & \nodata & 313$\degr$$\pm$3$\degr$\\ 			
MKO $J$\tablenotemark{a} & 11.53$\pm$0.04 & 11.22$\pm$0.04 & $-$0.31$\pm$0.05 \\
MKO $H$ & 10.37$\pm$0.04 & 10.39$\pm$0.04 &  $-$0.02$\pm$0.06 \\
MKO $K$ & 9.44$\pm$0.07 & 9.73$\pm$0.09 &  0.29$\pm$0.16 \\
$M_J$ & 15.00$\pm$0.17 & 14.69$\pm$0.17 & \nodata \\
$J-K$ & 2.08$\pm$0.08 & 1.49$\pm$0.10 & \nodata \\
\enddata
\tablenotetext{a}{Uncertainty-weighted average from FIRE ($-$0.30$\pm$0.05) and SpeX ($-$0.34$\pm$0.14) analyses.}
\end{deluxetable}

\section{Discussion}

\begin{figure*}
\epsscale{1.0}
\plottwo{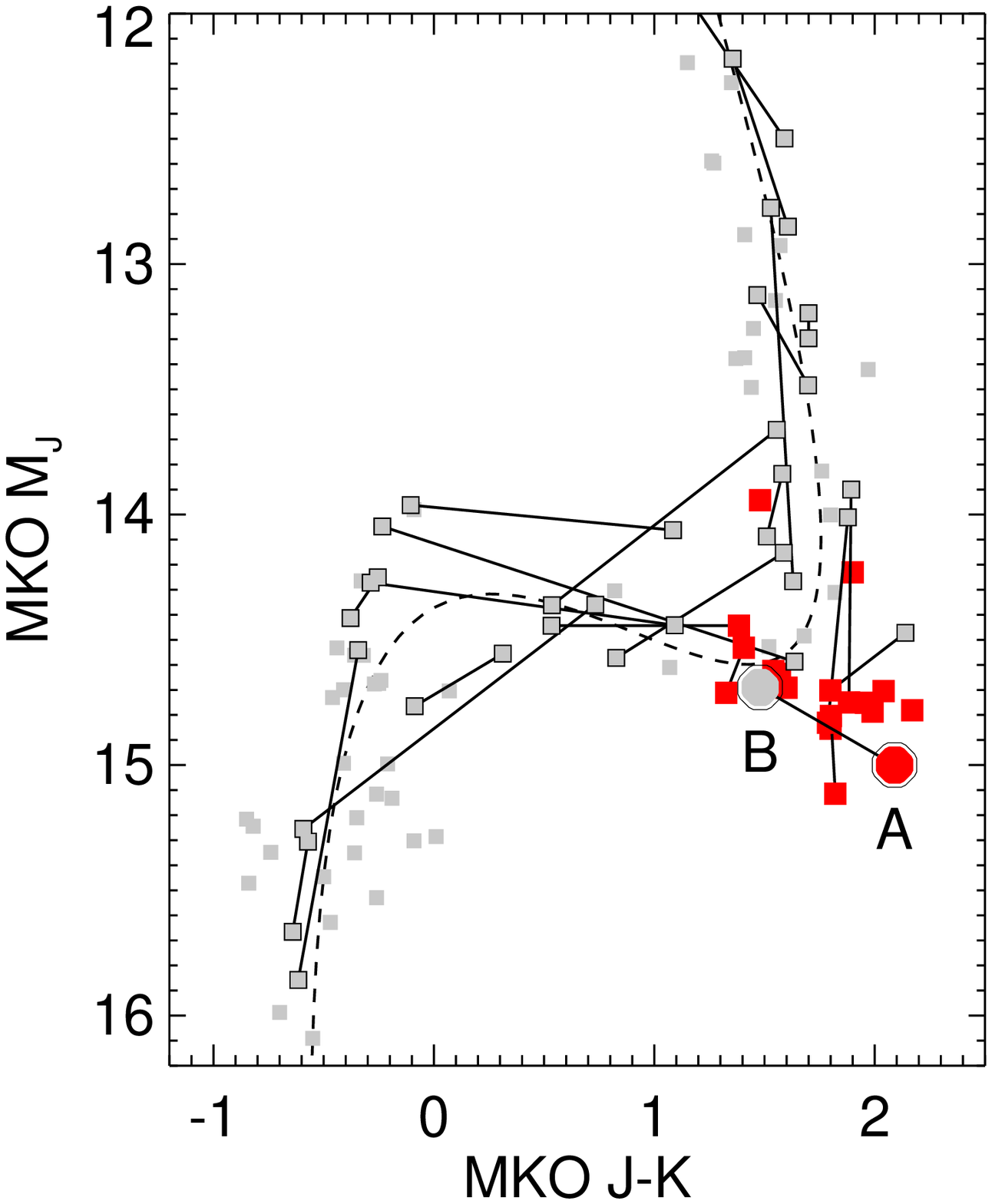}{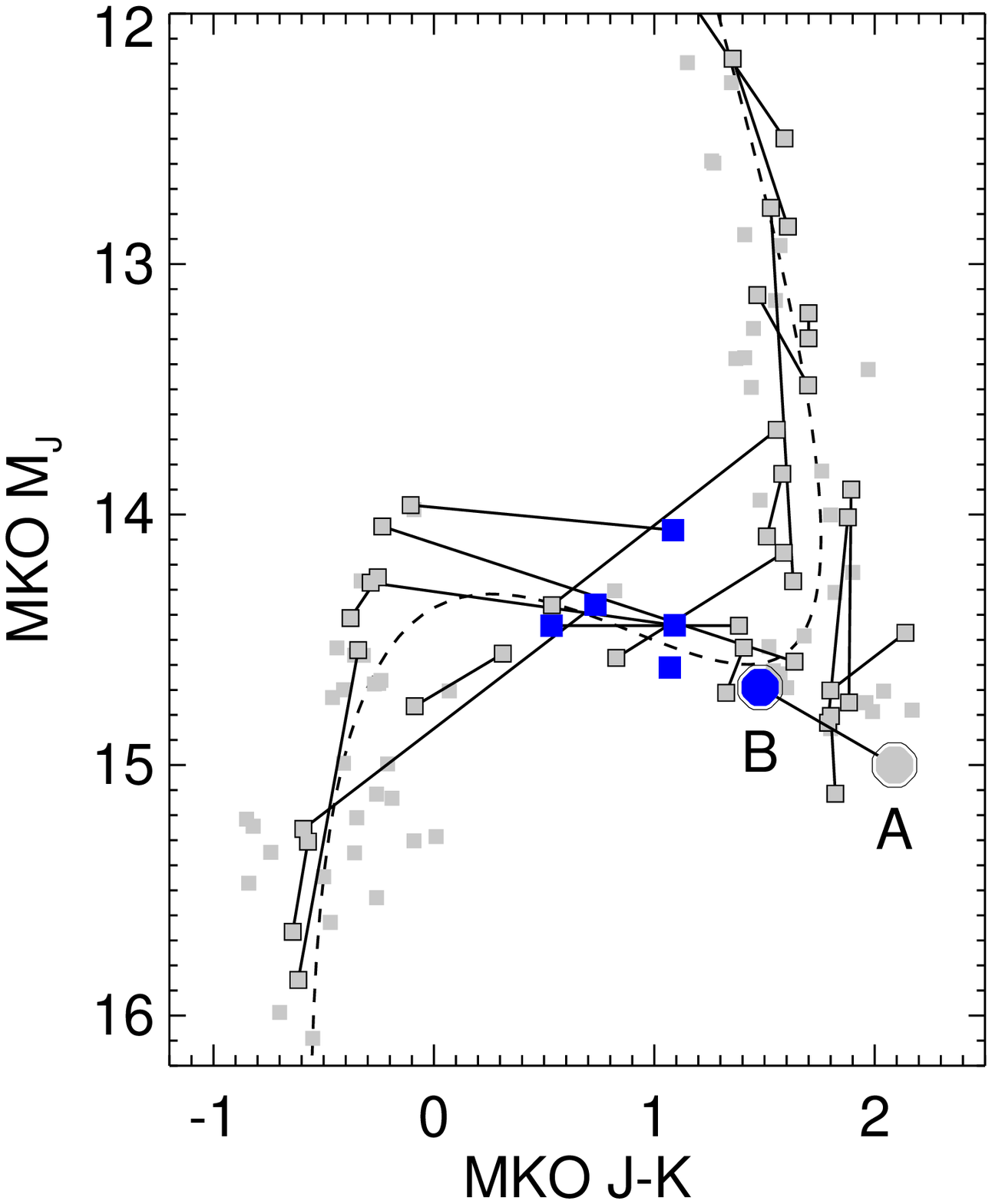} \\
\plottwo{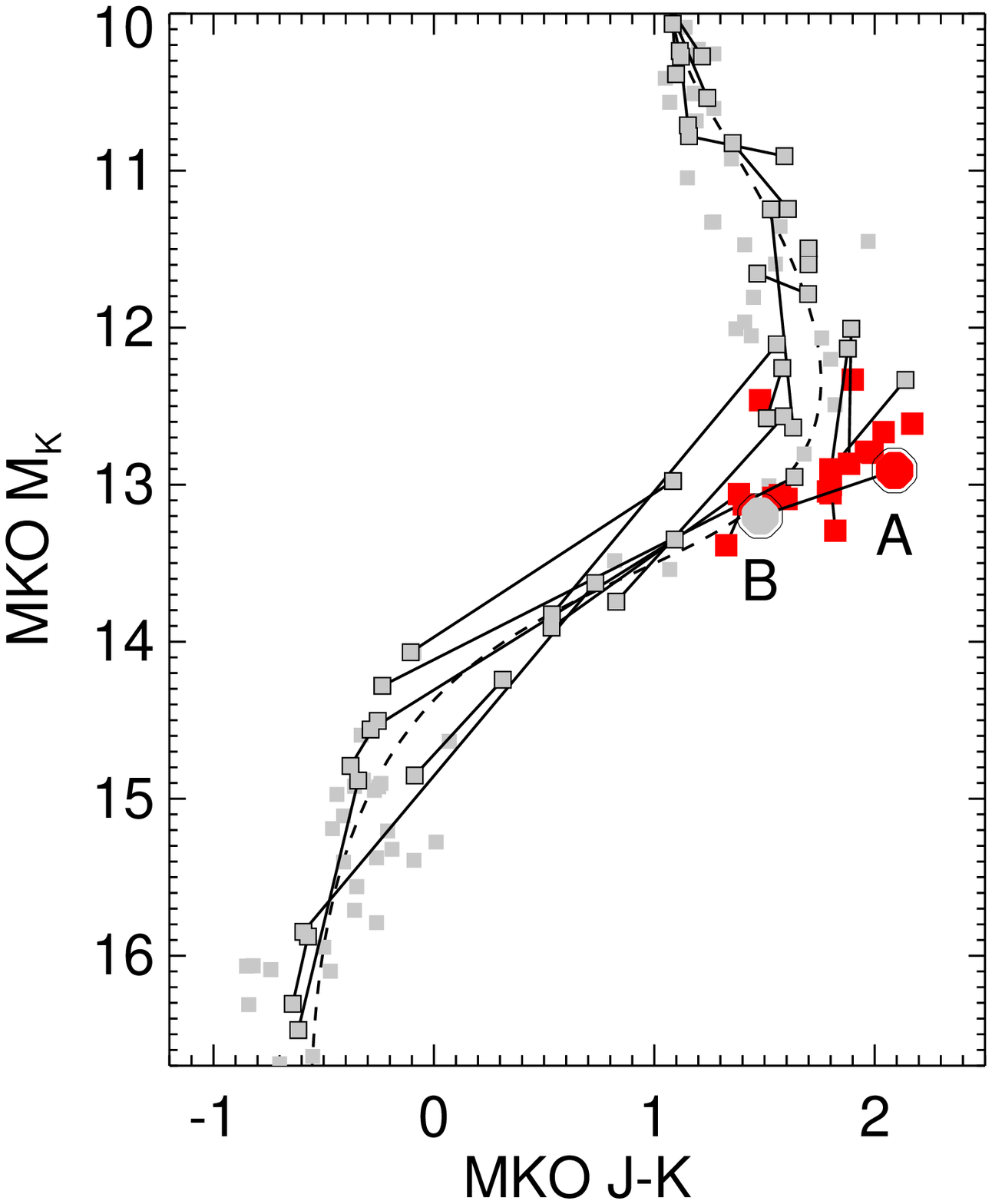}{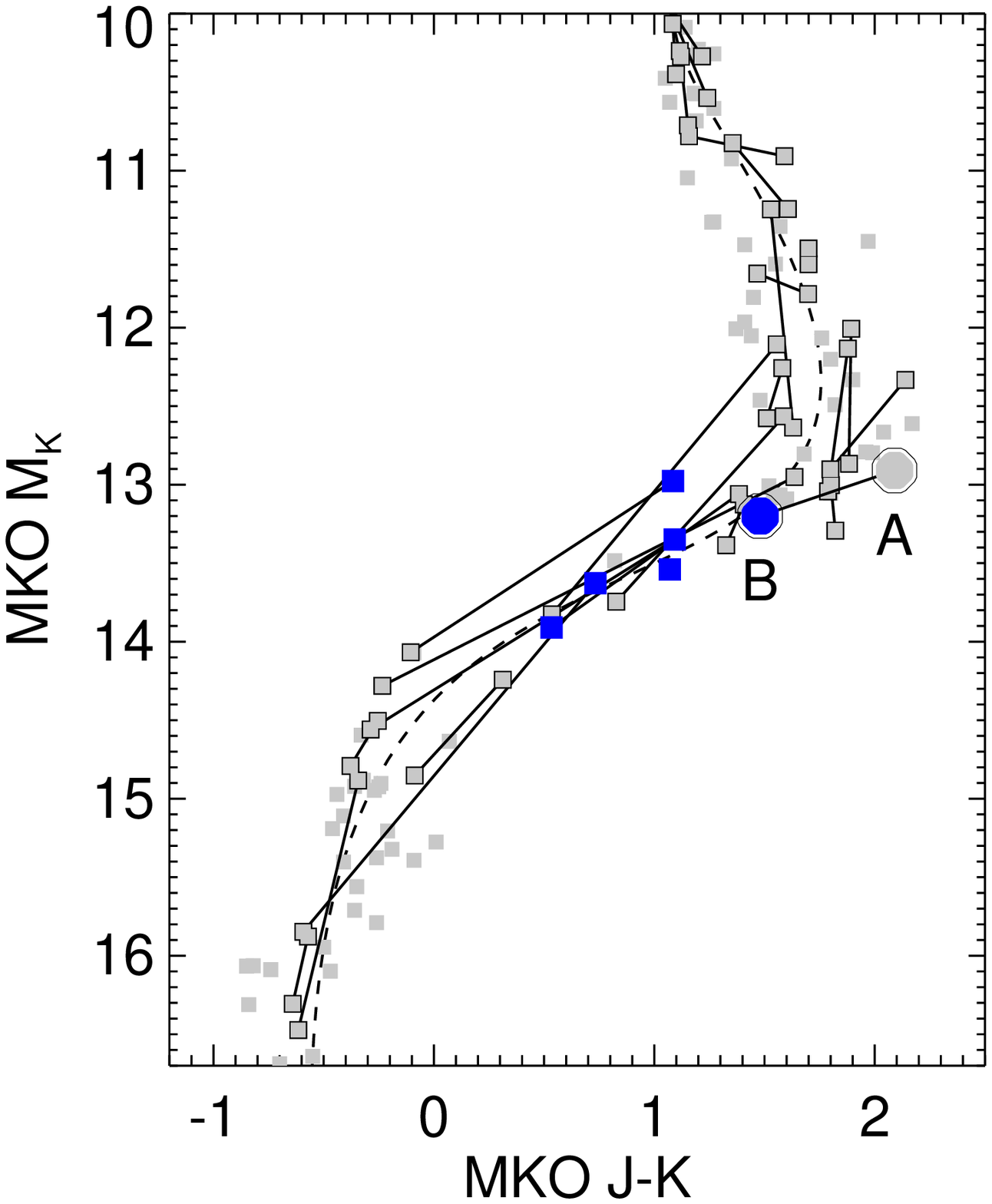}
\caption{Near-infrared color-magnitude diagrams (top:  $M_J$ versus $J-K$; bottom:  $M_K$ versus $J-K$) for field L and T dwarfs with reported parallax measurements having absolute magnitude and color uncertainties $\leq$0.3~mag.  Subdwarfs and young brown dwarfs are excluded. Components of binary systems (including {\namesh}AB) are indicated by outlined symbols and connected by solid lines.  L7-L8 dwarfs are highlighted in the left panels; T0-T1 dwarfs in the right panels.  {\namesh}A and B are somewhat redder and fainter than their equivalently classified counterparts.  Mean color-magnitude trends from \citet{2012ApJS..201...19D} are delineated by dashed lines.
}
\label{fig:cmd}
\end{figure*}

Our relative photometry allows us to infer individual component magnitudes and colors for {\namesh}AB and place them on color-magnitude diagrams, as shown in Figure~\ref{fig:cmd}.  As in our comparison with the spectral standards (Figure~\ref{fig:standards}), we see that both components are relatively red and subluminous for their spectral types, with {\namesh}A in particular being one of the reddest L dwarfs with a well-determined parallax.  
Unusually red $J-K$ colors for L dwarfs have been attributed to youth \citep{2006ApJ...639.1120K}, unusually thick clouds \citep{2008ApJ...686..528L} or both \citep{2011ApJ...729..128C,2013AJ....145....2F}.  The kinematics of {\namesh}AB appear to rule out membership in any of the known local young associations with the possible exception of the 40~Myr Argus association \citep{2013arXiv1303.5345M,2011ApJ...732...61Z}.  
However, other than the Li~I line in the optical spectrum of the primary, which sets a maximum age for the system of $\sim$4.5~Gyr (assuming M$_1$ $<$ 0.065~{\msun}, {\teff} $\approx$ 1350~K and evolutionary models from \citealt{2003A&A...402..701B}), there are no obvious indicators of low surface gravity in the low-resolution spectra of these sources.  We therefore conclude that thick clouds is the more likely interpretation for the red colors.  

Thick clouds may explain why {\namesh}AB is a flux reversal binary. Excess cloud opacity in the primary suppresses its 1~$\micron$ flux, whereas reduced cloud opacity in the secondary allows light to emerge from deeper, hotter layers at these wavelengths, making this component relatively brighter. 
This argument has been made to explain the relative brightnesses of other flux-reversal binaries, but this is the first time resolved spectroscopy has been reported for such a system.  Cloud-induced variations in the degree of flux reversal have also been previously suggested among candidate L/T transition spectral binaries \citep{2010ApJ...710.1142B}.

Why would there be a difference in the cloud properties of these coeval brown dwarfs?   Overall color trends among L and T dwarfs suggest that this binary may simply straddle the evolutionary stage during which mineral clouds are disrupted, and the presumably lower-mass and lower-temperature secondary has begun to lose its clouds first.  Note that the mass differential between the components could potentially constrain the timescale over which this evolutionary phase occurs \citep{2007ApJ...659..655B}.  An alternate---or possibly concurrent---hypothesis is that these two brown dwarfs have different cloud properties due to other factors; e.g., differences in surface gravity, rotation rate or viewing orientation (metallicity or age differences are assumed negligible).  In this case, the thermal evolution of the cloudy primary may have slowed due to its greater condensate opacity \citep{2008ApJ...689.1327S}, potentially leading to a situation in which the lower-temperature secondary is the {\it more} massive component of the system.  

These hypotheses are potentially testable for this system, as its relatively tight separation of (3.13$\pm$0.25~AU) makes it a promising (and rare) target for individual component mass measurements through astrometric and radial velocity (RV) orbital mapping.  Assuming a true semi-major axis of roughly 4~AU
and equal component masses of 0.04--0.065~{\msun} (appropriate for {\teff} $\approx$1300--1350~K and ages of 1--4.5~Gyr; \citealt{2009ApJ...702..154S,2003A&A...402..701B}), this system has a likely orbital period estimate of 22--28~yr, implying astrometric orbital motion of up to 0$\farcs$3 per year and in-plane RV variability amplitudes of 0.7--0.8~{\kms} (corresponding to RV differences of up to 1.6~{\kms}).  The RV velocity measurements are potentially feasible with current instrumentation \citep{2010ApJ...713..410B,2010ApJ...723..684B}, and the relatively dense stellar field around {\namesh}AB means that astrometric motion should be easily measurable over the course of a single year.

\acknowledgments

The authors thank Hugo Rivera at Magellan
and Eric Volquardsen at IRTF for their assistance with the observations.
K.L. acknowledges support from grant NNX12AI47G from the NASA Astrophysics Data Analysis Program.
The Center for Exoplanets and Habitable Worlds is supported by the
Pennsylvania State University, the Eberly College of Science, and the
Pennsylvania Space Grant Consortium.

Facilities: \facility{Magellan: FIRE}, \facility{IRTF: SpeX}


\begin{thebibliography}{56}
\expandafter\ifx\csname natexlab\endcsname\relax\def\natexlab#1{#1}\fi

\bibitem[{{Ackerman} \& {Marley}(2001)}]{2001ApJ...556..872A}
{Ackerman}, A.~S., \& {Marley}, M.~S. 2001, \apj, 556, 872

\bibitem[{{Allard} {et~al.}(2001){Allard}, {Hauschildt}, {Alexander},
  {Tamanai}, \& {Schweitzer}}]{2001ApJ...556..357A}
{Allard}, F., {Hauschildt}, P.~H., {Alexander}, D.~R., {Tamanai}, A., \&
  {Schweitzer}, A. 2001, \apj, 556, 357

\bibitem[{{Baraffe} {et~al.}(2003){Baraffe}, {Chabrier}, {Barman}, {Allard}, \&
  {Hauschildt}}]{2003A&A...402..701B}
{Baraffe}, I., {Chabrier}, G., {Barman}, T.~S., {Allard}, F., \& {Hauschildt},
  P.~H. 2003, \aap, 402, 701

\bibitem[{{Barman} {et~al.}(2011){Barman}, {Macintosh}, {Konopacky}, \&
  {Marois}}]{2011ApJ...733...65B}
{Barman}, T.~S., {Macintosh}, B., {Konopacky}, Q.~M., \& {Marois}, C. 2011,
  \apj, 733, 65

\bibitem[{{Bean} {et~al.}(2010){Bean}, {Seifahrt}, {Hartman}, {Nilsson},
  {Wiedemann}, {Reiners}, {Dreizler}, \& {Henry}}]{2010ApJ...713..410B}
{Bean}, J.~L., {Seifahrt}, A., {Hartman}, H., {Nilsson}, H., {Wiedemann}, G.,
  {Reiners}, A., {Dreizler}, S., \& {Henry}, T.~J. 2010, \apj, 713, 410

\bibitem[{{Blake} {et~al.}(2010){Blake}, {Charbonneau}, \&
  {White}}]{2010ApJ...723..684B}
{Blake}, C.~H., {Charbonneau}, D., \& {White}, R.~J. 2010, \apj, 723, 684

\bibitem[{{Burgasser}(2007)}]{2007ApJ...659..655B}
{Burgasser}, A.~J. 2007, \apj, 659, 655

\bibitem[{{Burgasser} {et~al.}(2010{\natexlab{a}}){Burgasser}, {Cruz},
  {Cushing}, {Gelino}, {Looper}, {Faherty}, {Kirkpatrick}, \&
  {Reid}}]{2010ApJ...710.1142B}
{Burgasser}, A.~J., {Cruz}, K.~L., {Cushing}, M., {Gelino}, C.~R., {Looper},
  D.~L., {Faherty}, J.~K., {Kirkpatrick}, J.~D., \& {Reid}, I.~N.
  2010{\natexlab{a}}, \apj, 710, 1142

\bibitem[{{Burgasser} {et~al.}(2006{\natexlab{a}}){Burgasser}, {Geballe},
  {Leggett}, {Kirkpatrick}, \& {Golimowski}}]{2006ApJ...637.1067B}
{Burgasser}, A.~J., {Geballe}, T.~R., {Leggett}, S.~K., {Kirkpatrick}, J.~D.,
  \& {Golimowski}, D.~A. 2006{\natexlab{a}}, \apj, 637, 1067

\bibitem[{{Burgasser} {et~al.}(2006{\natexlab{b}}){Burgasser}, {Kirkpatrick},
  {Cruz}, {Reid}, {Leggett}, {Liebert}, {Burrows}, \&
  {Brown}}]{2006ApJS..166..585B}
{Burgasser}, A.~J., {Kirkpatrick}, J.~D., {Cruz}, K.~L., {Reid}, I.~N.,
  {Leggett}, S.~K., {Liebert}, J., {Burrows}, A., \& {Brown}, M.~E.
  2006{\natexlab{b}}, \apjs, 166, 585

\bibitem[{{Burgasser} {et~al.}(2002){Burgasser}, {Marley}, {Ackerman},
  {Saumon}, {Lodders}, {Dahn}, {Harris}, \&
  {Kirkpatrick}}]{2002ApJ...571L.151B}
{Burgasser}, A.~J., {Marley}, M.~S., {Ackerman}, A.~S., {Saumon}, D.,
  {Lodders}, K., {Dahn}, C.~C., {Harris}, H.~C., \& {Kirkpatrick}, J.~D. 2002,
  \apjl, 571, L151

\bibitem[{{Burgasser} {et~al.}(2007){Burgasser}, {Reid}, {Siegler}, {Close},
  {Allen}, {Lowrance}, \& {Gizis}}]{2007prpl.conf..427B}
{Burgasser}, A.~J., {Reid}, I.~N., {Siegler}, N., {Close}, L., {Allen}, P.,
  {Lowrance}, P., \& {Gizis}, J. 2007, Protostars and Planets V, 427

\bibitem[{{Burgasser} {et~al.}(2010{\natexlab{b}}){Burgasser}, {Simcoe},
  {Bochanski}, {Saumon}, {Mamajek}, {Cushing}, {Marley}, {McMurtry}, {Pipher},
  \& {Forrest}}]{2010ApJ...725.1405B}
{Burgasser}, A.~J., {Simcoe}, R.~A., {Bochanski}, J.~J., {Saumon}, D.,
  {Mamajek}, E.~E., {Cushing}, M.~C., {Marley}, M.~S., {McMurtry}, C.,
  {Pipher}, J.~L., \& {Forrest}, W.~J. 2010{\natexlab{b}}, \apj, 725, 1405

\bibitem[{{Burrows} {et~al.}(2000){Burrows}, {Marley}, \&
  {Sharp}}]{2000ApJ...531..438B}
{Burrows}, A., {Marley}, M.~S., \& {Sharp}, C.~M. 2000, \apj, 531, 438

\bibitem[{{Burrows} \& {Sharp}(1999)}]{1999ApJ...512..843B}
{Burrows}, A., \& {Sharp}, C.~M. 1999, \apj, 512, 843

\bibitem[{{Burrows} {et~al.}(2006){Burrows}, {Sudarsky}, \&
  {Hubeny}}]{2006ApJ...640.1063B}
{Burrows}, A., {Sudarsky}, D., \& {Hubeny}, I. 2006, \apj, 640, 1063

\bibitem[{{Currie} {et~al.}(2011){Currie}, {Burrows}, {Itoh}, {Matsumura},
  {Fukagawa}, {Apai}, {Madhusudhan}, {Hinz}, {Rodigas}, {Kasper}, {Pyo}, \&
  {Ogino}}]{2011ApJ...729..128C}
{Currie}, T., {Burrows}, A., {Itoh}, Y., {Matsumura}, S., {Fukagawa}, M.,
  {Apai}, D., {Madhusudhan}, N., {Hinz}, P.~M., {Rodigas}, T.~J., {Kasper}, M.,
  {Pyo}, T., \& {Ogino}, S. 2011, \apj, 729, 128

\bibitem[{{Cushing} {et~al.}(2004){Cushing}, {Vacca}, \&
  {Rayner}}]{2004PASP..116..362C}
{Cushing}, M.~C., {Vacca}, W.~D., \& {Rayner}, J.~T. 2004, \pasp, 116, 362

\bibitem[{{Dupuy} \& {Liu}(2012)}]{2012ApJS..201...19D}
{Dupuy}, T.~J., \& {Liu}, M.~C. 2012, \apjs, 201, 19

\bibitem[{{Dupuy} {et~al.}(2009){Dupuy}, {Liu}, \&
  {Ireland}}]{2009ApJ...692..729D}
{Dupuy}, T.~J., {Liu}, M.~C., \& {Ireland}, M.~J. 2009, \apj, 692, 729

\bibitem[{{Faherty} {et~al.}(2013){Faherty}, {Rice}, {Cruz}, {Mamajek}, \&
  {N{\'u}{\~n}ez}}]{2013AJ....145....2F}
{Faherty}, J.~K., {Rice}, E.~L., {Cruz}, K.~L., {Mamajek}, E.~E., \&
  {N{\'u}{\~n}ez}, A. 2013, \aj, 145, 2

\bibitem[{{Geballe} {et~al.}(2002){Geballe}, {Knapp}, {Leggett}, {Fan},
  {Golimowski}, {Anderson}, {Brinkmann}, {Csabai}, {Gunn}, {Hawley},
  {Hennessy}, {Henry}, {Hill}, {Hindsley}, {Ivezi{\'c}}, {Lupton}, {McDaniel},
  {Munn}, {Narayanan}, {Peng}, {Pier}, {Rockosi}, {Schneider}, {Smith},
  {Strauss}, {Tsvetanov}, {Uomoto}, {York}, \& {Zheng}}]{2002ApJ...564..466G}
{Geballe}, T.~R., {Knapp}, G.~R., {Leggett}, S.~K., {Fan}, X., {Golimowski},
  D.~A., {Anderson}, S., {Brinkmann}, J., {Csabai}, I., {Gunn}, J.~E.,
  {Hawley}, S.~L., {Hennessy}, G., {Henry}, T.~J., {Hill}, G.~J., {Hindsley},
  R.~B., {Ivezi{\'c}}, {\v Z}., {Lupton}, R.~H., {McDaniel}, A., {Munn}, J.~A.,
  {Narayanan}, V.~K., {Peng}, E., {Pier}, J.~R., {Rockosi}, C.~M., {Schneider},
  D.~P., {Smith}, J.~A., {Strauss}, M.~A., {Tsvetanov}, Z.~I., {Uomoto}, A.,
  {York}, D.~G., \& {Zheng}, W. 2002, \apj, 564, 466

\bibitem[{{Hayashi} \& {Nakano}(1963)}]{1963PThPh..30..460H}
{Hayashi}, C., \& {Nakano}, T. 1963, Progress of Theoretical Physics, 30, 460

\bibitem[{{Kelson}(2003)}]{2003PASP..115..688K}
{Kelson}, D.~D. 2003, \pasp, 115, 688

\bibitem[{{Kirkpatrick}(2005)}]{2005ARA&A..43..195K}
{Kirkpatrick}, J.~D. 2005, \araa, 43, 195

\bibitem[{{Kirkpatrick} {et~al.}(2006){Kirkpatrick}, {Barman}, {Burgasser},
  {McGovern}, {McLean}, {Tinney}, \& {Lowrance}}]{2006ApJ...639.1120K}
{Kirkpatrick}, J.~D., {Barman}, T.~S., {Burgasser}, A.~J., {McGovern}, M.~R.,
  {McLean}, I.~S., {Tinney}, C.~G., \& {Lowrance}, P.~J. 2006, \apj, 639, 1120

\bibitem[{{Kirkpatrick} {et~al.}(2010){Kirkpatrick}, {Looper}, {Burgasser},
  {Schurr}, {Cutri}, {Cushing}, {Cruz}, {Sweet}, {Knapp}, {Barman},
  {Bochanski}, {Roellig}, {McLean}, {McGovern}, \&
  {Rice}}]{2010ApJS..190..100K}
{Kirkpatrick}, J.~D., {Looper}, D.~L., {Burgasser}, A.~J., {Schurr}, S.~D.,
  {Cutri}, R.~M., {Cushing}, M.~C., {Cruz}, K.~L., {Sweet}, A.~C., {Knapp},
  G.~R., {Barman}, T.~S., {Bochanski}, J.~J., {Roellig}, T.~L., {McLean},
  I.~S., {McGovern}, M.~R., \& {Rice}, E.~L. 2010, \apjs, 190, 100

\bibitem[{{Kirkpatrick} {et~al.}(1999){Kirkpatrick}, {Reid}, {Liebert},
  {Cutri}, {Nelson}, {Beichman}, {Dahn}, {Monet}, {Gizis}, \&
  {Skrutskie}}]{1999ApJ...519..802K}
{Kirkpatrick}, J.~D., {Reid}, I.~N., {Liebert}, J., {Cutri}, R.~M., {Nelson},
  B., {Beichman}, C.~A., {Dahn}, C.~C., {Monet}, D.~G., {Gizis}, J.~E., \&
  {Skrutskie}, M.~F. 1999, \apj, 519, 802

\bibitem[{{Knapp} {et~al.}(2004){Knapp}, {Leggett}, {Fan}, {Marley}, {Geballe},
  {Golimowski}, {Finkbeiner}, {Gunn}, {Hennawi}, {Ivezi{\'c}}, {Lupton},
  {Schlegel}, {Strauss}, {Tsvetanov}, {Chiu}, {Hoversten}, {Glazebrook},
  {Zheng}, {Hendrickson}, {Williams}, {Uomoto}, {Vrba}, {Henden}, {Luginbuhl},
  {Guetter}, {Munn}, {Canzian}, {Schneider}, \&
  {Brinkmann}}]{2004AJ....127.3553K}
{Knapp}, G.~R., {Leggett}, S.~K., {Fan}, X., {Marley}, M.~S., {Geballe}, T.~R.,
  {Golimowski}, D.~A., {Finkbeiner}, D., {Gunn}, J.~E., {Hennawi}, J.,
  {Ivezi{\'c}}, Z., {Lupton}, R.~H., {Schlegel}, D.~J., {Strauss}, M.~A.,
  {Tsvetanov}, Z.~I., {Chiu}, K., {Hoversten}, E.~A., {Glazebrook}, K.,
  {Zheng}, W., {Hendrickson}, M., {Williams}, C.~C., {Uomoto}, A., {Vrba},
  F.~J., {Henden}, A.~A., {Luginbuhl}, C.~B., {Guetter}, H.~H., {Munn}, J.~A.,
  {Canzian}, B., {Schneider}, D.~P., \& {Brinkmann}, J. 2004, \aj, 127, 3553

\bibitem[{{Konopacky} {et~al.}(2010){Konopacky}, {Ghez}, {Barman}, {Rice},
  {Bailey}, {White}, {McLean}, \& {Duch{\^e}ne}}]{2010ApJ...711.1087K}
{Konopacky}, Q.~M., {Ghez}, A.~M., {Barman}, T.~S., {Rice}, E.~L., {Bailey},
  J.~I., {White}, R.~J., {McLean}, I.~S., \& {Duch{\^e}ne}, G. 2010, \apj, 711,
  1087

\bibitem[{{Kumar}(1962)}]{1962AJ.....67S.579K}
{Kumar}, S.~S. 1962, \aj, 67, 579

\bibitem[{{Lane} {et~al.}(2001){Lane}, {Zapatero Osorio}, {Britton},
  {Mart{\'{\i}}n}, \& {Kulkarni}}]{2001ApJ...560..390L}
{Lane}, B.~F., {Zapatero Osorio}, M.~R., {Britton}, M.~C., {Mart{\'{\i}}n},
  E.~L., \& {Kulkarni}, S.~R. 2001, \apj, 560, 390

\bibitem[{{Leggett} {et~al.}(2000){Leggett}, {Geballe}, {Fan}, {Schneider},
  {Gunn}, {Lupton}, {Knapp}, {Strauss}, {McDaniel}, {Golimowski}, {Henry},
  {Peng}, {Tsvetanov}, {Uomoto}, {Zheng}, {Hill}, {Ramsey}, {Anderson},
  {Annis}, {Bahcall}, {Brinkmann}, {Chen}, {Csabai}, {Fukugita}, {Hennessy},
  {Hindsley}, {Ivezi{\'c}}, {Lamb}, {Munn}, {Pier}, {Schlegel}, {Smith},
  {Stoughton}, {Thakar}, \& {York}}]{2000ApJ...536L..35L}
{Leggett}, S.~K., {Geballe}, T.~R., {Fan}, X., {Schneider}, D.~P., {Gunn},
  J.~E., {Lupton}, R.~H., {Knapp}, G.~R., {Strauss}, M.~A., {McDaniel}, A.,
  {Golimowski}, D.~A., {Henry}, T.~J., {Peng}, E., {Tsvetanov}, Z.~I.,
  {Uomoto}, A., {Zheng}, W., {Hill}, G.~J., {Ramsey}, L.~W., {Anderson}, S.~F.,
  {Annis}, J.~A., {Bahcall}, N.~A., {Brinkmann}, J., {Chen}, B., {Csabai}, I.,
  {Fukugita}, M., {Hennessy}, G.~S., {Hindsley}, R.~B., {Ivezi{\'c}}, {\v Z}.,
  {Lamb}, D.~Q., {Munn}, J.~A., {Pier}, J.~R., {Schlegel}, D.~J., {Smith},
  J.~A., {Stoughton}, C., {Thakar}, A.~R., \& {York}, D.~G. 2000, \apjl, 536,
  L35

\bibitem[{{Liu} {et~al.}(2006){Liu}, {Leggett}, {Golimowski}, {Chiu}, {Fan},
  {Geballe}, {Schneider}, \& {Brinkmann}}]{2006ApJ...647.1393L}
{Liu}, M.~C., {Leggett}, S.~K., {Golimowski}, D.~A., {Chiu}, K., {Fan}, X.,
  {Geballe}, T.~R., {Schneider}, D.~P., \& {Brinkmann}, J. 2006, \apj, 647,
  1393

\bibitem[{{Looper} {et~al.}(2008{\natexlab{a}}){Looper}, {Gelino}, {Burgasser},
  \& {Kirkpatrick}}]{2008ApJ...685.1183L}
{Looper}, D.~L., {Gelino}, C.~R., {Burgasser}, A.~J., \& {Kirkpatrick}, J.~D.
  2008{\natexlab{a}}, \apj, 685, 1183

\bibitem[{{Looper} {et~al.}(2008{\natexlab{b}}){Looper}, {Kirkpatrick},
  {Cutri}, {Barman}, {Burgasser}, {Cushing}, {Roellig}, {McGovern}, {McLean},
  {Rice}, {Swift}, \& {Schurr}}]{2008ApJ...686..528L}
{Looper}, D.~L., {Kirkpatrick}, J.~D., {Cutri}, R.~M., {Barman}, T.,
  {Burgasser}, A.~J., {Cushing}, M.~C., {Roellig}, T., {McGovern}, M.~R.,
  {McLean}, I.~S., {Rice}, E., {Swift}, B.~J., \& {Schurr}, S.~D.
  2008{\natexlab{b}}, \apj, 686, 528

\bibitem[{{Luhman}(2013)}]{2013ApJ...767L...1L}
{Luhman}, K.~L. 2013, \apjl, 767, L1

\bibitem[{{Lunine} {et~al.}(1989){Lunine}, {Hubbard}, {Burrows}, {Wang}, \&
  {Garlow}}]{1989ApJ...338..314L}
{Lunine}, J.~I., {Hubbard}, W.~B., {Burrows}, A., {Wang}, Y.-P., \& {Garlow},
  K. 1989, \apj, 338, 314

\bibitem[{{Magazzu} {et~al.}(1993){Magazzu}, {Martin}, \&
  {Rebolo}}]{1993ApJ...404L..17M}
{Magazzu}, A., {Martin}, E.~L., \& {Rebolo}, R. 1993, \apjl, 404, L17

\bibitem[{{Mamajek}(2013)}]{2013arXiv1303.5345M}
{Mamajek}, E.~E. 2013, ArXiv e-prints

\bibitem[{{Marley} {et~al.}(2012){Marley}, {Saumon}, {Cushing}, {Ackerman},
  {Fortney}, \& {Freedman}}]{2012ApJ...754..135M}
{Marley}, M.~S., {Saumon}, D., {Cushing}, M., {Ackerman}, A.~S., {Fortney},
  J.~J., \& {Freedman}, R. 2012, \apj, 754, 135

\bibitem[{{Morley} {et~al.}(2012){Morley}, {Fortney}, {Marley}, {Visscher},
  {Saumon}, \& {Leggett}}]{2012ApJ...756..172M}
{Morley}, C.~V., {Fortney}, J.~J., {Marley}, M.~S., {Visscher}, C., {Saumon},
  D., \& {Leggett}, S.~K. 2012, \apj, 756, 172

\bibitem[{{Radigan} {et~al.}(2012){Radigan}, {Jayawardhana}, {Lafreni{\`e}re},
  {Artigau}, {Marley}, \& {Saumon}}]{2012ApJ...750..105R}
{Radigan}, J., {Jayawardhana}, R., {Lafreni{\`e}re}, D., {Artigau}, {\'E}.,
  {Marley}, M., \& {Saumon}, D. 2012, \apj, 750, 105

\bibitem[{{Rayner} {et~al.}(2003){Rayner}, {Toomey}, {Onaka}, {Denault},
  {Stahlberger}, {Vacca}, {Cushing}, \& {Wang}}]{2003PASP..115..362R}
{Rayner}, J.~T., {Toomey}, D.~W., {Onaka}, P.~M., {Denault}, A.~J.,
  {Stahlberger}, W.~E., {Vacca}, W.~D., {Cushing}, M.~C., \& {Wang}, S. 2003,
  \pasp, 115, 362

\bibitem[{{Rebolo} {et~al.}(1992){Rebolo}, {Martin}, \&
  {Magazzu}}]{1992ApJ...389L..83R}
{Rebolo}, R., {Martin}, E.~L., \& {Magazzu}, A. 1992, \apjl, 389, L83

\bibitem[{{Saumon} \& {Marley}(2008)}]{2008ApJ...689.1327S}
{Saumon}, D., \& {Marley}, M.~S. 2008, \apj, 689, 1327

\bibitem[{{Simcoe} {et~al.}(2008){Simcoe}, {Burgasser}, {Bernstein}, {Bigelow},
  {Fishner}, {Forrest}, {McMurtry}, {Pipher}, {Schechter}, \&
  {Smith}}]{2008SPIE.7014E..27S}
{Simcoe}, R.~A., {Burgasser}, A.~J., {Bernstein}, R.~A., {Bigelow}, B.~C.,
  {Fishner}, J., {Forrest}, W.~J., {McMurtry}, C., {Pipher}, J.~L.,
  {Schechter}, P.~L., \& {Smith}, M. 2008, in Society of Photo-Optical
  Instrumentation Engineers (SPIE) Conference Series, Vol. 7014, Society of
  Photo-Optical Instrumentation Engineers (SPIE) Conference Series

\bibitem[{{Simcoe} {et~al.}(2010){Simcoe}, {Burgasser}, {Bochanski},
  {Schechter}, {Bernstein}, {Bigelow}, {Pipher}, {Forrest}, {McMurtry},
  {Smith}, \& {Fishner}}]{2010SPIE.7735E..38S}
{Simcoe}, R.~A., {Burgasser}, A.~J., {Bochanski}, J.~J., {Schechter}, P.~L.,
  {Bernstein}, R.~A., {Bigelow}, B.~C., {Pipher}, J.~L., {Forrest}, W.,
  {McMurtry}, C., {Smith}, M.~J., \& {Fishner}, J. 2010, in Society of
  Photo-Optical Instrumentation Engineers (SPIE) Conference Series, Vol. 7735,
  Society of Photo-Optical Instrumentation Engineers (SPIE) Conference Series

\bibitem[{{Simons} \& {Tokunaga}(2002)}]{2002PASP..114..169S}
{Simons}, D.~A., \& {Tokunaga}, A. 2002, \pasp, 114, 169

\bibitem[{{Skemer} {et~al.}(2012){Skemer}, {Hinz}, {Esposito}, {Burrows},
  {Leisenring}, {Skrutskie}, {Desidera}, {Mesa}, {Arcidiacono}, {Mannucci},
  {Rodigas}, {Close}, {McCarthy}, {Kulesa}, {Agapito}, {Apai}, {Argomedo},
  {Bailey}, {Boutsia}, {Briguglio}, {Brusa}, {Busoni}, {Claudi}, {Eisner},
  {Fini}, {Follette}, {Garnavich}, {Gratton}, {Guerra}, {Hill}, {Hoffmann},
  {Jones}, {Krejny}, {Males}, {Masciadri}, {Meyer}, {Miller}, {Morzinski},
  {Nelson}, {Pinna}, {Puglisi}, {Quanz}, {Quiros-Pacheco}, {Riccardi},
  {Stefanini}, {Vaitheeswaran}, {Wilson}, \& {Xompero}}]{2012ApJ...753...14S}
{Skemer}, A.~J., {Hinz}, P.~M., {Esposito}, S., {Burrows}, A., {Leisenring},
  J., {Skrutskie}, M., {Desidera}, S., {Mesa}, D., {Arcidiacono}, C.,
  {Mannucci}, F., {Rodigas}, T.~J., {Close}, L., {McCarthy}, D., {Kulesa}, C.,
  {Agapito}, G., {Apai}, D., {Argomedo}, J., {Bailey}, V., {Boutsia}, K.,
  {Briguglio}, R., {Brusa}, G., {Busoni}, L., {Claudi}, R., {Eisner}, J.,
  {Fini}, L., {Follette}, K.~B., {Garnavich}, P., {Gratton}, R., {Guerra},
  J.~C., {Hill}, J.~M., {Hoffmann}, W.~F., {Jones}, T., {Krejny}, M., {Males},
  J., {Masciadri}, E., {Meyer}, M.~R., {Miller}, D.~L., {Morzinski}, K.,
  {Nelson}, M., {Pinna}, E., {Puglisi}, A., {Quanz}, S.~P., {Quiros-Pacheco},
  F., {Riccardi}, A., {Stefanini}, P., {Vaitheeswaran}, V., {Wilson}, J.~C., \&
  {Xompero}, M. 2012, \apj, 753, 14

\bibitem[{{Stephens} {et~al.}(2009){Stephens}, {Leggett}, {Cushing}, {Marley},
  {Saumon}, {Geballe}, {Golimowski}, {Fan}, \& {Noll}}]{2009ApJ...702..154S}
{Stephens}, D.~C., {Leggett}, S.~K., {Cushing}, M.~C., {Marley}, M.~S.,
  {Saumon}, D., {Geballe}, T.~R., {Golimowski}, D.~A., {Fan}, X., \& {Noll},
  K.~S. 2009, \apj, 702, 154

\bibitem[{{Tinney} {et~al.}(2003){Tinney}, {Burgasser}, \&
  {Kirkpatrick}}]{2003AJ....126..975T}
{Tinney}, C.~G., {Burgasser}, A.~J., \& {Kirkpatrick}, J.~D. 2003, \aj, 126,
  975

\bibitem[{{Tokunaga} {et~al.}(2002){Tokunaga}, {Simons}, \&
  {Vacca}}]{2002PASP..114..180T}
{Tokunaga}, A.~T., {Simons}, D.~A., \& {Vacca}, W.~D. 2002, \pasp, 114, 180

\bibitem[{{Tsuji} {et~al.}(1996){Tsuji}, {Ohnaka}, \&
  {Aoki}}]{1996A&A...305L...1T}
{Tsuji}, T., {Ohnaka}, K., \& {Aoki}, W. 1996, \aap, 305, L1+

\bibitem[{{Vacca} {et~al.}(2003){Vacca}, {Cushing}, \&
  {Rayner}}]{2003PASP..115..389V}
{Vacca}, W.~D., {Cushing}, M.~C., \& {Rayner}, J.~T. 2003, \pasp, 115, 389

\bibitem[{{Zuckerman} {et~al.}(2011){Zuckerman}, {Rhee}, {Song}, \&
  {Bessell}}]{2011ApJ...732...61Z}
{Zuckerman}, B., {Rhee}, J.~H., {Song}, I., \& {Bessell}, M.~S. 2011, \apj,
  732, 61

\end{thebibliography}

\end{document}